\documentclass[draftclsnofoot, onecolumn, 12pt]{IEEEtran}
\IEEEoverridecommandlockouts
\usepackage{cite}
\usepackage{graphicx}
\usepackage{psfrag}
\usepackage{subfigure}
\usepackage{amsmath}
\usepackage{amssymb}
\usepackage{epsfig}
\usepackage{color}
\usepackage{epsf}
\usepackage{algorithmic}
\usepackage{algorithm}
\usepackage{epsfig}
\usepackage{fancyhdr}
\usepackage{setspace}
\usepackage{amsfonts}
\usepackage{times}
\usepackage{latexsym}
\usepackage{dsfont}
\usepackage{amssymb}
\usepackage{amsmath}
\usepackage{cite}
\usepackage{verbatim}
\usepackage{subfigure}

\newtheorem{assumption}{Assumption}

\newtheorem{lemma}{Lemma}

\newtheorem{proposition}{Proposition}

\def\bb0{{\mathbb{0}}}

\def\ba{{\mathbf{a}}}
\def\bb{{\mathbf{b}}}

\def\bff{{\mathbf{f}}}

\def\bs{{\mathbf{s}}}

\def\bv{{\mathbf{v}}}
\def\bw{{\mathbf{w}}}
\def\bx{{\mathbf{x}}}
\def\by{{\mathbf{y}}}

\def\b0{{\mathbf{0}}}

\def\bA{{\mathbf{A}}}

\def\bF{{\mathbf{F}}}
\def\bG{{\mathbf{G}}}
\def\bH{{\mathbf{H}}}
\def\bI{{\mathbf{I}}}

\def\bV{{\mathbf{V}}}
\def\bW{{\mathbf{W}}}
\def\bX{{\mathbf{X}}}
\def\bY{{\mathbf{Y}}}

\def\sf0{{\mathsf{0}}}

\def\Nt{{N_\mathrm{T}}}         %
\def\Nr{{N_\mathrm{R}}}         %
\def\Tohd{{T_\mathrm{OHD}}}     %
\def\Tframe{T_\mathrm{frame}}   %
\def\rhoeff{\rho_\mathrm{eff}}  %
\def\Rsum{\bar{R}_\mathrm{sum}}
\def\fd{{f_\mathrm{D}}}         %
\def\taup{{\tau_\mathrm{p}}}    %
\def\taut{{\tau_\mathrm{t}}}    %
\def\tauf{{\tau_\mathrm{f}}}    %
\def\Pf{{P_\mathrm{F}}}         %
\def\dRsum{{\dot{R}_\mathrm{sum}}}  
\def\ddRsum{{\ddot{R}_\mathrm{sum}}} 
\def\sigmaH{\sigma^2_{\widetilde{\bH}}} 
\def\Reff{\bar{R}_{\mathrm{eff}}}

\begin{document}

\title{On the Overhead of Interference Alignment: Training, Feedback, and Cooperation\thanks{The authors at The University of Texas were supported by the Office of Naval Research (ONR) under grant N000141010337. The work of A. Lozano is supported by the FET FP7 Project
265578 ``HIATUS''.}
}

\author{\IEEEauthorblockN{Omar El Ayach, Angel Lozano, and Robert W. Heath, Jr.\footnote{O. El Ayach and R.W. Heath, Jr are with The University of Texas at Austin, Austin, TX 78712 USA (e-mail: \{omarayach, rheath\}@mail.utexas.edu). A. Lozano is with Universitat Pompeu Fabra, Barcelona, Spain (e-mail: angel.lozano@upf.edu)}\\}}

\maketitle

\begin{abstract}

Interference alignment (IA) is a cooperative transmission strategy that, under some conditions, achieves the interference channel's maximum number of degrees of freedom. Realizing IA gains, however, is contingent upon providing transmitters with sufficiently accurate channel knowledge. In this paper, we study the performance of IA in multiple-input multiple-output systems where channel knowledge is acquired through training and analog feedback. We design the training and feedback system to maximize IA's effective sum-rate: a non-asymptotic performance metric that accounts for estimation error, training and feedback overhead, and channel selectivity. We characterize effective sum-rate with overhead in relation to various parameters such as signal-to-noise ratio, Doppler spread, and feedback channel quality. A main insight from our analysis is that, by properly designing the CSI acquisition process, IA can provide good sum-rate performance in a very wide range of fading scenarios. Another observation from our work is that such overhead-aware analysis can help solve a number of practical network design problems. To demonstrate the concept of overhead-aware network design, we consider the example problem of finding the optimal number of cooperative IA users based on signal power and mobility.

\end{abstract}


\section{Introduction}

Interference alignment (IA) for the multiple-input multiple-output (MIMO) interference channel is a cooperative transmission strategy that attempts to structure interfering signals such that they occupy a reduced dimensional space when observed at the receivers~\cite{Cadambe2008, GuoJafar}. Alignment often enables achieving the maximum number of degrees of freedom (DoF)~\cite{Cadambe2008, GuoJafar}. Precoding transmitted signals to carefully align them at the receivers, however, requires knowledge of the interfering channels in the system, collectively known as channel state information (CSI). Perfect CSI is assumed to be available when designing most IA algorithms \cite{Cadambe2008,Peters2010, Gomadam2008, Tresch_achievability} or reporting genie-aided IA gains. Practical systems, however, acquire receiver CSI with the help of training sequences or pilots~\cite{goldsmith2005wireless}. Such CSI can then be shared with the transmitters via feedback. As a result, practical CSI is imperfect and comes with an overhead signaling cost, both of which penalize the effective data rates achieved. Realizing the gains of IA is therefore contingent upon providing systems with sufficiently accurate CSI at a manageable overhead cost.

Several approaches have been proposed to fulfill IA's transmit CSI requirement~\cite{Thukral2009,Krishnamachari2009,Ayach2010}, typically assuming perfect CSI at the receiver. The feedback strategy in \cite{Thukral2009} proposes to use Grassmannian codebooks to compress and improve CSI feedback in single-antenna frequency extended IA systems. The feedback strategy was then extended to multiantenna frequency extended systems in~\cite{Krishnamachari2009}. Both~\cite{Thukral2009} and~\cite{Krishnamachari2009} guarantee that limited feedback preserves the number of DoF by scaling the number of feedback bits with SNR, thus making codebooks prohibitively large~\cite{Caire_shamai}. To overcome the problem of scaling codebook size, and relax the reliance on frequency selectivity for quantization, \cite{Ayach2010} proposed an analog feedback strategy for constant MIMO interference channels. Using analog feedback, a constant data rate gap from perfect CSI performance was shown, as long as the SNRs on the forward and feedback links are order-wise equal. A limitation of the analysis in~\cite{Thukral2009,Krishnamachari2009, Ayach2010}, however, is that the number of DoF remains the primary performance metric considered. IA's sum-rate performance at finite SNR, especially when accounting for the time spent on overhead signaling, has yet to be considered.

Attempts to more directly analyze or reduce overhead are limited to~\cite{nosratinia_IAtraining, Ayach2011_GDC, Peters2010a, guillaud2011interference}. To analyze the effect of overhead, \cite{nosratinia_IAtraining} considers the effective number of spatial DoF of an IA system with training and feedback. By considering DoF, however, \cite{nosratinia_IAtraining} implicitly characterizes performance at infinitely high SNR. Alternatively, \cite{Ayach2011_GDC} reduces codebook size to limit overhead in limited feedback IA systems by leveraging temporal correlation without providing any overhead-aware analysis. In another line of work, information about the network topology is used to partition users into optimally sized alignment groups~\cite{Peters2010a}. In~\cite{guillaud2011interference}, IA is applied to partially connected interference channels. User grouping and partial connectivity, however, only reduces the number of channels that must be shared without suggesting an efficient training and feedback strategy. 

In this paper, we characterize the performance of a MIMO IA system that is designed for perfect CSI operation yet only has access to imperfect CSI through training and analog feedback~\cite{Ayach2010, Marzetta2006, caire710multiuser}. Thus, the performance demonstrated in this paper constitutes a lower bound for systems that are designed to be more robust to imperfect CSI through improved precoding strategies such as~\cite{GuillaudDeterministic} for example. We adopt a block-fading model wherein the channel remains constant over the block length, and varies independently across blocks. In contrast with earlier work on IA with feedback, we precisely model channel selectivity by leveraging the relationship between block-fading and continuous-fading channels shown in~\cite{jindal2010unified}. This relationship allows us to define the concept of Doppler spread in a block fading channel and explicitly relate the size of the coherence block to that Doppler spread. Since both CSI acquisition and data transmission must now occur within the limits of a single coherence block, the IA system is faced with a non-trivial tradeoff: too much overhead leaves little time for payload data transmission, whereas too little overhead results in large sum-rate losses due to poor CSI quality~\cite{jindal2010unified, hassibi_training, kobayashi2009training, lozano2008interplay, santipach2010optimization}. In this paper, we design the training and analog feedback system to maximize IA's \emph{effective sum-rate}, a non-asymptotic performance metric that accounts for both CSI quality and CSI acquisition overhead. CSI acquisition overhead is a fundamental concept that was largely neglected in earlier work on IA with imperfect CSI. 

We begin by giving a tractable expression for the IA sum-rate in genie-aided systems with perfect CSI, and extend the analysis under a general model for imperfect CSI. We then specialize our results to a system with training and analog feedback by characterizing CSI quality as a function of system parameters such as training overhead, feedback overhead and transmit power on both forward and reverse links. This results in a tractable expression for IA's effective sum-rate, which we proceed to optimize. To give a closed-form solution for the optimal effective sum-rate, we build on the method in~\cite{jindal2010unified} and optimize a series expansion of the objective function. Initial results were reported in our previous work \cite{Ayach2011_asilomar}. In this paper, we complete IA's performance analysis by analytically characterizing its maximum achievable effective sum-rate and the corresponding optimum overhead budget. The main insights and conclusions that can be drawn from the effective sum-rate analysis can be summarized as follows:
\begin{itemize}
\item  Practical IA performance is not only a function of basic system parameters such as network size and SNR, but is tightly related to quantities such as Doppler spread, and feedback channel quality. Moreover, the dependence of both the maximum effective sum-rate, and the corresponding optimal overhead budget, on the various system parameters can be characterized accurately.
\item By properly designing the training and feedback stages, IA can be made both feasible and beneficial in a wide range of fading scenarios, even when its relatively high overhead is considered.
\item Overhead-aware analysis is essential to the design of IA networks. As an example of this observation, we use the overhead analysis to give simple results on the optimal number of cooperative IA users for channels with varying levels of selectivity.
\end{itemize}


Throughout this paper, we use the following notation: $\bA$ is a matrix; $\ba$ is a vector; $a$ is a scalar; $(\cdot)^*$ denotes the conjugate transpose; $\|\ba\|$ denotes the $2$-norm of $\ba$; $\left|a\right|$ is the absolute value of $a$;  $\bI_N$ is the $N \times N$ identity matrix; $\mathcal{CN}(\ba,\bA)$ is a complex Gaussian random vector with mean $\ba$ and covariance matrix $\bA$; $(a_1, \hdots, a_k)$ is an ordered set; $\mathbb{E}\left[\cdot\right]$ denotes expectation.

\section{System Model} \label{sec:background}

\begin{figure}
 \centering
	\includegraphics[width=5in]{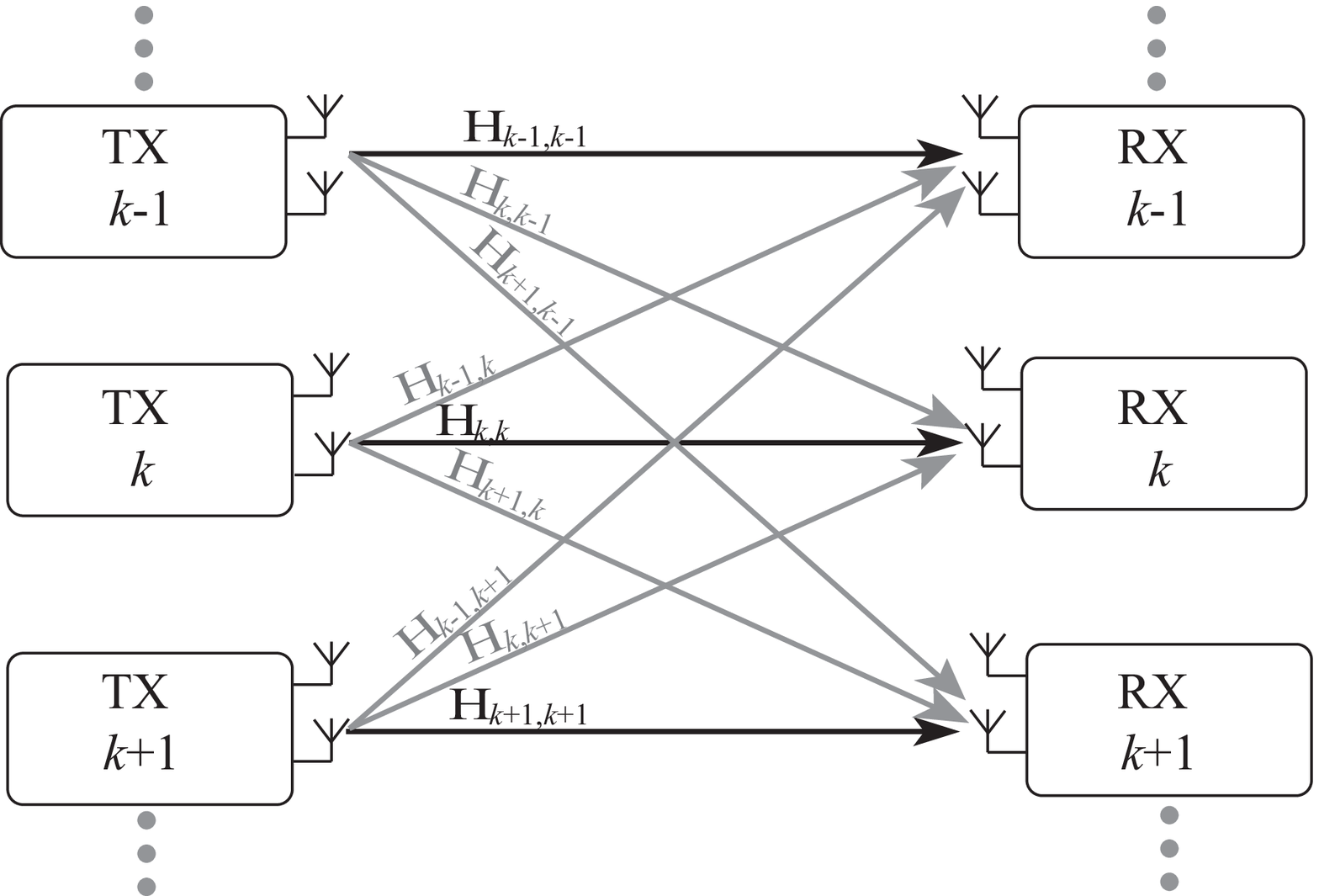}
	\caption{$K$-User MIMO interference channel model}
	\label{fig:system_model}
\end{figure}

Consider the $K$-user narrowband MIMO interference channel shown in Fig. \ref{fig:system_model} in which transmitter $i$ communicates with its paired receiver $i$ and interferes with all other receivers, $\ell \neq i$. For simplicity of exposition, consider a homogeneous network where all transmitters are equipped with $\Nt$ antennas and all receivers with $\Nr$ antennas, and each node pair communicates via $d \leq \min(\Nt,\Nr)$ independent spatial streams. The results can be generalized to a different number of streams or antennas at each node, provided that IA remains feasible~\cite{razaviyayn_DoF}.


Assuming perfect time and frequency synchronization, the sampled baseband signal at receiver $i$ can be written as
\begin{equation}
\mathbf{y}_{i}= \sqrt{\frac{P}{d}}\mathbf{H}_{i,i}\mathbf{F}_{i}\mathbf{s}_{i} + \sum_{\ell \neq i} \sqrt{\frac{P}{d}}\mathbf{H}_{i,\ell}\mathbf{F}_{\ell}\mathbf{s}_{\ell} + \mathbf{v}_{i},
\label{eqn:narrowsig_model}
\end{equation}
where $\mathbf{y}_{i}$ is the $\Nr \times 1$ received signal vector, $P$ is the transmit power, $\mathbf{H}_{i,\ell}$ is the $\Nr \times \Nt$ discrete-time effective baseband channel matrix from transmitter $\ell$ to receiver $i$, $\mathbf{F}_{i}=\left[\bff_i^{1},\ \hdots,\ \bff_{i}^{d}\right]$ is transmitter $i$'s $\Nt \times d$ precoding matrix, $\bs_{i}$ is the $d \times 1$ transmitted symbol vector at node $i$ such that $\mathbb{E}\left[\bs_i\bs_i^*\right]=\bI_d$, and $\mathbf{v}_{i}$ is a vector of i.i.d complex Gaussian noise samples with covariance matrix $\sigma^2\mathbf{I}_{\Nr}$. The channels $\mathbf{H}_{i,\ell}$ are assumed to be independent across users and each with i.i.d $\mathcal{CN}(0,1)$ entries. Large-scale fading can be included in the system model at the expense of a more involved exposition in Section \ref{sec:analog_feedback}. 

The received signal at \emph{transmitter} $i$ on the feedback channel is
\begin{equation}
\overleftarrow{\mathbf{y}}_{i}=\sqrt{\frac{\Pf}{\Nr}}\bG_{i,i}\overleftarrow{\bx}_{i} + \sum_{\ell \neq i} \sqrt{\frac{\Pf}{\Nr}}\bG_{\ell,i}\overleftarrow{\bx}_{\ell} + \overleftarrow{\mathbf{v}}_{i},
\label{eqn:reverse_model}
\end{equation}
where $\Pf$ is the feedback power available such that $\Pf/P=\gamma$, $\bG_{\ell,i}$ is the $\Nt \times \Nr$ discrete time feedback channel between receiver $\ell$ and transmitter $i$ with i.i.d $\mathcal{CN}(0,1)$ entries, $\overleftarrow{\bx}_i$ is the symbol vector with unit variance entries sent by receiver $i$, and $\overleftarrow{\mathbf{v}}_{i}$ is a complex vector of i.i.d circularly symmetric white Gaussian noise with covariance matrix $\sigma^2 \mathbf{I_{\Nt}}$. The forward and feedback channels are assumed to be independent in the error analysis of Section \ref{sec:analog_feedback}, i.e., a frequency division duplexed system or a general non-reciprocal system is assumed.

We adopt a block-fading channel model in which channels remain fixed for a period, $\Tframe$, but vary independently from block to block. To model the effect of channel selectivity on IA performance, we set the block length to $\Tframe=\frac{1}{2\fd}$, where $\fd$ plays the role of the block fading channel's \emph{effective Doppler spread}. The definition of $\fd$ is motivated by the results in \cite{jindal2010unified} showing a relationship between continuous fading and block fading systems. 
To enable IA over such a channel, both CSI acquisition and payload data transmission must occur within the coherence time $\Tframe $, or else the CSI acquired becomes obsolete. The IA system then encounters a well-known tension between CSI acquisition and data transmission~\cite{kobayashi2009training, lozano2008interplay, jindal2010unified, hassibi_training, santipach2010optimization}, and must allocate resources to each of the processes to optimize overall performance.

\begin{figure}
 \centering
	\includegraphics[width=4in]{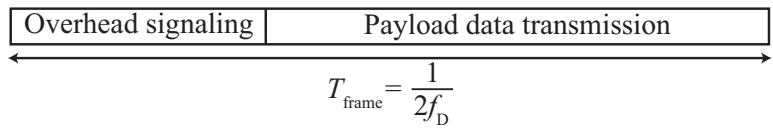}
	\caption{The overhead model adopted in which training and feedback consume resources that would otherwise be used for data transmission.}
	\label{fig:overhead_model}
\end{figure}

To account for CSI acquisition overhead, and to accurately characterize the \emph{effective data rate} achieved by IA, we adopt the overhead model shown in Fig. \ref{fig:overhead_model}. In this model, overhead signaling consumes time resources that could otherwise be used for data transmission, i.e., CSI acquisition penalizes effective sum-rate. For such an overhead model, the effective sum-rate (in bits/s/Hz) can be written as~\cite{kobayashi2009training, jindal2010unified, hassibi_training}
\begin{equation}
\Reff\left(P,\Tohd \right)=\left(\frac{\Tframe -\Tohd }{\Tframe }\right)\Rsum (P,\Tohd )
\label{eqn:throughput_simple}
\end{equation}
where $\Tohd $ is the total time spent on training and feeding back channels, and $\Rsum (P, \Tohd )$ is the average sum-rate in bits/s/Hz achieved by IA on the channel uses allocated for payload transmission. Using (\ref{eqn:throughput_simple}), and previous insights into IA performance, we highlight the tradeoff between overhead signaling and data transmission. Increasing overhead improves CSI quality and in turn improves $\Rsum (P, \Tohd )$, but the relative period over which $\Rsum (P, \Tohd )$ can be achieved shrinks. A similar tension exists when lowering overhead; less overhead allows more channel uses for data transmission but the sum-rate per channel use suffers due to poor CSI quality. The objective then becomes maximizing the effective sum rate given in (\ref{eqn:throughput_simple}) by optimally trading off overhead with data transmission~\cite{kobayashi2009training, lozano2008interplay, jindal2010unified, hassibi_training, santipach2010optimization}. Throughout this paper, we treat $\Rsum (P, \Tohd )$ as an information-theoretic quantity, and thus derive mutual information-based sum-rates achievable without errors. IA performance can also be analyzed from the perspective of fixed-rate transmission where metrics such as bit error rate may be of interest~\cite{IA_BER}.

\section{Interference Alignment: An Average Sum-Rate Analysis} \label{sec:IArates}

This section derives the average sum-rate achieved by IA in both genie-aided networks where channels are known perfectly, as well as practical systems where CSI is imperfect.

\subsection{Interference Alignment with Perfect CSI} \label{sec:perfectCSI}

IA often achieves the full number of DoF supported by MIMO interference channels. In cases where the full DoF cannot be guaranteed, IA has been shown to provide significant gains in high-SNR sum-rate~\cite{Peters2010,Ayach2009, Gomadam2008}. While this paper focuses on IA, even better performance could be achieved with other precoding algorithms that seek a balance between interference minimization and signal power maximization~\cite{Peters2010, MMSE-IA,Luo_WMMSE}. The algorithms in~\cite{Peters2010, MMSE-IA,Luo_WMMSE}, however, do not readily lend themselves to average sum-rate analysis.


To analyze IA sum-rates, we begin by examining the effective channels created after precoding and combining. For tractability, we focus on IA with a simple per-stream zero-forcing (ZF) receiver. Recall that in the high (but finite) SNR regime, where IA is most useful, gains from more involved receiver designs are limited. In such a system, receiver $i$ projects its signal onto the columns of the zero-forcing combiner $\bW_i=\left[\bw_i^1,\ \hdots,\ \bw_i^m,\ \hdots,\ \bw_i^d\right]$ which gives
\begin{align}
\begin{split}
(\bw_i^m)^*\by_i= & \sqrt{\frac{P}{d}}(\bw_i^m)^*\mathbf{H}_{i,i}\bff_{i}^{m} s_{i}^{m}+\sqrt{\frac{P}{d}}\sum\limits_{(k,\ell) \neq (i,m)}(\bw_i^m)^*\mathbf{H}_{i,k}\bff_{k}^{\ell} s_{k}^{\ell}+(\bw_i^m)^*\bv_i.
\label{eqn:recv_proj}
\end{split}
\end{align}
At the output of these linear receivers $\bw_i^m$, the conditions for perfect IA can be stated as \cite{Gomadam2008}
\begin{eqnarray}
(\bw_i^m)^*\mathbf{H}_{i,k}\bff_{k}^{\ell}=0, &\quad& \forall (k,\ell) \neq (i,m) \label{eqn:conditions1}\\
\left|(\bw_i^m)^*\mathbf{H}_{i,i}\bff_{i}^{m}\right|\geq c >0, &\quad& \forall i,m, \label{eqn:conditions2}
\label{eqn:conditions}
\end{eqnarray}
where alignment is guaranteed by (\ref{eqn:conditions1}), and (\ref{eqn:conditions2}) is satisfied almost surely~\cite{Cadambe2008,Gomadam2008}. 

As a result of conditions (\ref{eqn:conditions1}) and (\ref{eqn:conditions2}), the combination of IA and ZF effectively creates $Kd$ non-interfering scalar channels. The maximum mutual information across these channels is achieved via Gaussian signaling which yields an instantaneous sum-rate given by 
\begin{equation}
R_\mathrm{sum}=\sum\limits_{i=1}^{K}\sum\limits_{m=1}^{d}\log_2\left(1+\frac{\frac{P}{d} \left|(\bw_i^m)^*\bH_{i,i}\bff_{i}^{m}\right|^2}{\sigma^2}\right).
\end{equation}
To derive an expression for the average sum-rate, i.e., $\Rsum=\mathbb{E}\left[R_\mathrm{sum}\right]$, we first give the following lemma.
\begin{lemma}[\!\it{\cite[Appendix A]{Ayach2010}}\it ] The effective direct channels $(\bw_i^m)^*\bH_{i,i}\bff_{i}^{m}$ are independent and Gaussian distributed with unit variance if: (i) the precoders $\bF_i$ are unitary and are generated by an IA solution that does not consider the direct channels $\bH_{i,i}$, and (ii) the combiners $\bW_i$ are calculated to simply zero-force inter-user and inter-stream interference.
\label{lemma:gaussian}
\end{lemma}
The conditions Lemma 1 places on precoder and combiner calculation are satisfied by most IA solutions such as \cite{Cadambe2008, Gomadam2008, Peters2010, Tresch_achievability}. Hence, as a result of Lemma \ref{lemma:gaussian}, the scalar point-to-point channels created by the combination of IA and ZF experience Rayleigh fading. As a result, the average sum-rate can be written in exponential integral form as~\cite{shin2003capacity,ozarow1994information}
\begin{equation}
\Rsum (\rho) =\sum\limits_{i=1}^{K}\sum\limits_{m=1}^{d} \mathbb{E}\left[\log_2\left(1+\frac{\frac{P}{d} \left|(\bw_i^m)^*\bH_{i,i}\bff_{i}^{m}\right|^2}{\sigma^2}\right)\right] = Kd\log_2(e)e^{1/ \rho}E_1\left(\frac{1}{\rho}\right),
\label{eqn:av_rate}
\end{equation}
which is written as a function of the per-stream SNR, $\rho=\frac{P}{d\sigma^2}$, and $E_1(\eta)=\int\limits_{1}^{\infty}t^{-1}e^{-\eta t}dt$ is an exponential integral. 

\subsection{Interference Alignment with CSI from Training and Feedback} \label{sec:imperfect_rates}

When the channels are not known perfectly, interference cannot be aligned perfectly. Misalignment leads to ``leakage interference'', which reduces the signal-to-interference-plus-noise ratio (SINR) in the desired signal space. Moreover, imperfect knowledge of the direct channel implies that receivers will perform mismatched decoding~\cite{lapidoth2002fading}, again reducing effective SINR. In this section, we examine the effect of imperfect CSI on the performance of an IA system that is optimized for perfect-CSI operation, i.e., a system that does not consider CSI imperfection in its design. Thus, the performance results demonstrated in this paper can be improved upon by adopting precoding algorithms that are more robust to CSI errors such as~\cite{GuillaudDeterministic}.

Consider an IA system in which transmitters use a common set of channel estimates as input to an IA solution such as \cite{Gomadam2008,Cadambe2008,Peters2010, Tresch_achievability}, i.e., they calculate imperfect IA precoders $\widehat{\bF}_i$ and combiners $\widehat{\bW}_i$. Denote the channel estimates as $\widehat{\bH}_{i,\ell}$ and the corresponding error as $\widetilde{\bH}_{i,\ell}=\bH_{i,\ell}-\widehat{\bH}_{i,\ell}$. In this system, the IA solution satisfies
\begin{eqnarray}
(\widehat{\bw}_i^m)^*\widehat{\bH}_{i,k}\widehat{\bff}_{k}^{\ell}=0, &\quad& \forall (k,\ell)\neq (i,m)  \label{eqn:conditions1hat}\\
\left|(\widehat{\bw}_i^m)^*\widehat{\bH}_{i,i}\widehat{\bff}_{i}^{m}\right|\geq c >0, &\quad& \forall i,m.\label{eqn:conditions2hat}
\end{eqnarray} 

We assume receivers obtain perfect knowledge of the combiners $\widehat{\bW}_i$ and the imperfect effective direct channels $\widehat{\bw}_i^m\widehat{\bH}_{i,i}\widehat{\bff}_i^m$ for detection, an assumption similar to~\cite{guillaud_cellular, kobayashi2009training, Krishnamachari2009, Thukral2009, Ayach2010}\footnote{In fact \cite{guillaud_cellular,Krishnamachari2009,Thukral2009} place a stronger assumption summarized by the receivers' knowledge of the exact imperfect CSI known to the transmitters. The two assumptions are functionally equivalent from the perspective of the sum-rate analysis, i.e., all that is needed is the receivers' knowledge of $\widehat{\bw}_i^m$ and of the scalars $\widehat{\bw}_i^m\widehat{\bH}_{i,i}\widehat{\bff}_i^m$.} whose relaxation is a topic of future work. In general, receiver side information about the effective channels can be acquired blindly~\cite{honigblind} or via additional training or silent phases~\cite{caire710multiuser}. For such an IA system, the received signal after projection is
\begin{align}
\begin{split}
(\widehat{\bw}_i^m)^*\by_i=\sqrt{\frac{P}{d}} & (\widehat{\bw}_i^m)^*\widehat{\bH}_{i,i}\widehat{\bff}_{i}^{m} s_{i}^{m} +\sqrt{\frac{P}{d}}\sum\limits_{k, \ell}(\widehat{\bw}_i^m)^*\widetilde{\bH}_{i,k}\widehat{\bff}_{k}^{\ell} s_{k}^{\ell}+(\widehat{\bw}_i^m)^*\bv_i,
\label{eqn:recv_proj_imp} 
\end{split}
\end{align}
where we have used the fact that conditions (\ref{eqn:conditions1hat}) and (\ref{eqn:conditions2hat}) are satisfied, thus $(\widehat{\bw}_i^m)^*\bH_{i,k}\widehat{\bff}_{k}^{\ell}= (\widehat{\bw}_i^m)^*(\widehat{\bH}_{i,k}+\widetilde{\bH}_{i,k})\widehat{\bff}_{k}^{\ell}= (\widehat{\bw}_i^m)^*\widetilde{\bH}_{i,k}\widehat{\bff}_{k}^{\ell}$. 

Analyzing the maximum sum-rates achievable on the channel in (\ref{eqn:recv_proj_imp}) is in general difficult, as it requires optimizing the distribution of the input symbols $\bs_i$ for the interference channel in (\ref{eqn:recv_proj_imp}). Recall, however, that our objective is to analyze a system optimized for \emph{perfect CSI operation}, i.e. one that does not account for CSI imperfection. This enables making the following assumptions that would be expected from a system optimized for perfect CSI operation. 

\begin{assumption}
Transmitters use a typical Gaussian codebook made up of i.i.d. symbols to form the symbol vectors $\bs_i$. Such a signaling codebook, which was optimal for the interference free channels created by IA with perfect CSI, may no longer be optimal now that CSI is imperfect. 
\label{as:codebook}
\end{assumption}
\begin{assumption}
Receivers perform nearest neighbor decoding using the estimates $\widehat{\bH}_{i,i}$. Nearest neighbor decoding would again be optimal with perfect CSI. The nearest neighbor decoder, the channel estimates and the signaling codebook together satisfy the conditions outlined in \cite{lapidoth2002fading} for Corollary 3.0.1 of \cite{lapidoth2002fading} to hold with equality, meaning that the estimation error plays the role of an additional source of Gaussian noise irrespective of its actual distribution. 
\label{as:decoding}
\end{assumption}

Under Assumptions \ref{as:codebook}-\ref{as:decoding}, and combining the results of \cite{lapidoth2002fading} and \cite{lapidoth1996nearest}, the average sum-rate achieved can be written as
\begin{equation}
\Rsum (\rho) =\sum\limits_{i=1}^{K}\sum\limits_{m=1}^{d} \mathbb{E}\left[\log_2\left(1+\frac{\frac{P}{d} \left|(\widehat{\bw}_i^m)^*\widehat{\bH}_{i,i}\widehat{\bff}_{i}^{m}\right|^2}{\sum\limits_{k,\ell}\mathbb{E}\left[\frac{P}{d} \left|(\widehat{\bw}_i^m)^*\widetilde{\bH}_{i,k}\widehat{\bff}_{k}^{\ell}\right|^2\right]+\sigma^2}\right)\right]
\label{eqn:av_rate_imperfect}
\end{equation}
where we note that the outer expectation is now only over the fading on the direct channel and not the interference. Therefore, the leakage interference terms $(\widehat{\bw}_i^m)^*\widetilde{\bH}_{i,k}\widehat{\bff}_{k}^{\ell}$ indeed play the role of independent sources of additive Gaussian noise, regardless of their distribution.

When the entries of $\widetilde{\bH}_{i,k}$ are zero-mean and uncorrelated with a variance of $\sigmaH$, it follows that $\mathbb{E}\left[\frac{P}{d} \left|(\widehat{\bw}_i^m)^*\widetilde{\bH}_{i,k}\widehat{\bff}_{k}^{\ell}\right|^2\right]=\frac{P}{d}\sigmaH$, thus the denominator in (\ref{eqn:av_rate_imperfect}) is simply $KP\sigma^2_{\widetilde{\bH}}+\sigma^2$. Moreover, if the estimates $\widehat{\bH}_{i,k}$ are MMSE estimates of $\bH_{i,k}$, the entries of $\widehat{\bH}_{i,k}$ have a variance of $1-\sigmaH$. This results in an effective average SINR that can be written as 
\begin{equation}
\rhoeff =\frac{\rho(1-\sigma^2_{\widetilde{\bH}})}{\rho Kd\sigma^2_{\widetilde{\bH}}+1},
\end{equation}
where $\rho$ is the per-stream SNR defined in (\ref{eqn:av_rate}). If the estimated direct channels $\widehat{\bH}_{i,i}$ is Gaussian, the average sum-rate achieved by IA with imperfect CSI is again given in exponential integral form as 
\begin{equation}
\Rsum (\rhoeff) = Kd\log_2(e)e^{1/ \rhoeff}E_1\left(\frac{1}{\rhoeff}\right).
\label{eqn:av_rate_final}
\end{equation}
To evaluate sum-rate achieved by IA, one must now characterize $\rhoeff$ or equivalently $\sigmaH$. In Section \ref{sec:analog_feedback} we specialize our result for a system with training and analog CSI feedback and later optimize IA's \textit{effective data rate with overhead} in Section \ref{sec:overhead_opt}.

\section{Training and Analog feedback} \label{sec:analog_feedback}

We propose to split the acquisition of CSI at the transmitter into three main phases. First, the transmitters train the forward channels via pilots. Second, the receivers train the feedback channels via pilots, setting the stage for the forward transmitters to estimate the feedback information in the next stage. Finally, the receivers feedback information about the forward channels in an analog fashion, i.e., as unquantized complex symbols. We can characterize the CSI error introduced in the CSI acquisition phase by examining the three stages.

\subsection{Forward and Feedback Channel Training} \label{sec:training}
In the first training phase, each transmitter $k$ sends an orthogonal pilot sequence matrix $\mathbf{\Phi}_k$, i.e., $\mathbf{\Phi}_i\mathbf{\Phi}_k^*=\delta_{ik}\bI_\Nt$, over a training period $\taut$~\cite{marzetta1999blast}. Pilot orthogonality imposes the constraint $\taut \geq K\Nt$. Each receiver $i$ then observes the $\Nr \times \taut$ matrix
\begin{equation}
\bY_i=\sqrt{\frac{\taut P}{\Nt}}\sum\limits_{k=1}^{K}\bH_{i,k}\mathbf{\Phi}_k+\bV_i, \quad \forall i, 
\label{eqn:training}
\end{equation}
where $\bV_i$ is an $\Nr \times \taut$ matrix of noise terms. Using $\bY_i$, receiver $i$ calculates an MMSE estimate of its incoming channels $\bH_{i,k}\ \forall k$ given by
\begin{equation}
\widehat{\bH}^{r}_{i,k}=\frac{\sqrt{\frac{\taut P}{\Nt}}}{\sigma^2+\frac{\taut P}{\Nt}}\bY_i\mathbf{\Phi}_k^*, \quad \forall k.
\end{equation} 
where the superscript $(\cdot)^r$ emphasizes that $\widehat{\bH}^{r}_{i,k}$ are the channel estimates gathered at the receiver before they are relayed back to the transmitters and further corrupted. At the output of this first training stage, the channel estimates $\widehat{\bH}^r_{i,k}$ have i.i.d. $\mathcal{CN}(0, \frac{\taut P/\Nt}{\sigma^2+\taut P/\Nt})$ entries with corresponding errors $\widetilde{\bH}^r_{i,k}\sim \mathcal{CN}(0, \frac{\sigma^2}{\sigma^2+\taut P/\Nt})$.

The feedback channel training phase proceeds similarly. Namely, the receivers transmit orthogonal pilot sequences over a training period $\taup \geq K\Nr$. The transmitters independently compute MMSE estimates of their incoming channels, resulting in estimates $\widehat{\bG}_{k,i}\sim \mathcal{CN}\left(0,\frac{\frac{\taup \Pf}{\Nr}}{\sigma^2+\frac{\taup \Pf}{\Nr}}\right)$ with corresponding error terms $\widetilde{\bG}_{k,i}\sim\mathcal{CN}\left(0,\frac{\sigma^2}{\sigma^2+\frac{\taup \Pf}{\Nr}}\right)$.

\subsection{Analog Feedback} \label{sec:feedback}

After forward and feedback channel training, the receivers feedback their channel estimates $\widehat{\bH}^r_{i,k}$ in an analog fashion during a feedback period $\tauf$. This is accomplished by first post-multiplying each $\Nr \times K\Nt$ feedback matrix $\left[\widehat{\bH}_{i,1}^r \ldots \widehat{\bH}_{i,K}^r\right]$ with a $K\Nt \times \tauf$ matrix $\mathbf{\Psi}_i$ such that $\mathbf{\Psi}_i\mathbf{\Psi}_k^*=\delta_{i,k}\bI_{K\Nt}$~\cite{Marzetta2006,Ayach2010}. The spreading matrices $\mathbf{\Psi}_i$ orthogonalize the feedback from different users and facilitate estimation. This orthogonality constraint requires that $\tauf \geq K^2\Nt$. The transmitted $\Nr \times \tauf$ feedback matrix $\overleftarrow{\bX}_i$ from receiver $i$ can be written as~\cite{Marzetta2006,Ayach2010}
\begin{equation}
\overleftarrow{\bX}_i=\sqrt{\frac{\tauf \Pf}{K\Nt\Nr}\left(\frac{\taut P/\Nt}{\sigma^2+\taut P/\Nt}\right)^{-1}}\left[\widehat{\bH}_{i,1}^r \ldots \widehat{\bH}_{i,K}^r\right]\mathbf{\Psi}_i, 
\end{equation}
where the leading scalar is to ensure that the average transmit power constraints are satisfied with equality, i.e., one can verify that $\mathbb{E}\left[\mathrm{trace}\left(\overleftarrow{\bX}_i \overleftarrow{\bX}_i^*\right)\right]=\tauf\Pf$.
We write the concatenated $K\Nt \times \tauf$ matrix of feedback symbols observed by all transmitters as 
\begin{equation}
\overleftarrow{\bY}_\mathrm{f}=\sqrt{\frac{\tauf \Pf}{K\Nt\Nr}\left(\frac{\taut P/\Nt}{\sigma^2+\taut P/\Nt}\right)^{-1}}\sum\limits_{i=1}^{K}\left[\begin{array}{c} \bG_{i,1} \\ \vdots \\ \bG_{i,K}\end{array}\right]\left[\widehat{\bH}_{i,1}^r \ldots \widehat{\bH}_{i,K}^r\right]\mathbf{\Psi}_i + \bV, 
\end{equation}
where $\bV$ is the $K\Nt \times \tauf$ matrix of i.i.d Gaussian noise.

To simplify the performance analysis, we make the same assumption as in \cite{Ayach2010}: at the end of the feedback phase, the transmitters cooperate by sharing their rows of the received feedback matrix $\overleftarrow{\bY}_\mathrm{f}$ which enables them to form a unified estimate of the forward channels $\bH_{i,k}$. We refer the reader to \cite{Ayach2010} for a discussion of this cooperative assumption and for alternative non-cooperative approaches that are shown to perform close to this special case.

Under this cooperative assumption, the transmitters estimate $\bH_{i,k}\ \forall k$ by first isolating the feedback sent by receiver $i$. They post-multiply their received symbols by $\mathbf{\Psi}_i^*$ to compute
\begin{equation}
\overleftarrow{\bY}_\mathrm{f}\mathbf{\Psi}_i^*=\sqrt{\frac{\tauf \Pf}{K\Nt\Nr}\left(\frac{\taut P/\Nt}{\sigma^2+\taut P/\Nt}\right)^{-1}}\underbrace{\left[\begin{array}{c} \bG_{i,1} \\ \vdots \\ \bG_{i,K}\end{array}\right]}_{\bG_i}\left[\widehat{\bH}_{i,1}^r \ldots \widehat{\bH}_{i,K}^r\right] + \bV\mathbf{\Psi}_i^*.
\label{eqn:Ypsi}
\end{equation}
The transmitters then compute a common linear MMSE estimate of the forward channels $\bH_{i,k}\ \forall i,k$ using their feedback channel estimates $\widehat{\bG}_{i,k}\ \forall i,k$, and assuming that $K\Nt \geq \Nr$ so that the estimation problem is well posed. After a lengthy yet standard application of the orthogonality principle and the matrix inversion lemma, the MMSE estimate is given by
\begin{equation}
\widehat{\bH}_i=\sqrt{\frac{K\Nt\Nr}{\tauf P_{f}}\left(\frac{\taut P/\Nt}{\sigma^2+\taut P/\Nt}\right)^{-1}}\left(\widehat{\bG}_i^*\widehat{\bG}_i+\gamma_1\widehat{\bG}_i^*\widehat{\bG}_i+\gamma_2 \bI_{\Nr}\right)^{-1}\widehat{\bG}_i^*\overleftarrow{\bY}_\mathrm{f}\mathbf{\Psi}_i^*,
\label{eqn:mmse_estimate}
\end{equation} 
where we have written (\ref{eqn:mmse_estimate}) in terms of $\widehat{\bH}_i=\left[\widehat{\bH}_{i,1},\ \hdots ,\ \widehat{\bH}_{i,K}\right]\ \forall i$, the concatenated estimate of the channels $\bH_i=\left[\bH_{i,1},\ \hdots,\ \bH_{i,K}\right]\ \forall i$, for the sake of notational brevity.
The constants $\gamma_1$ and $\gamma_2$ are the MMSE regularization factors. For completeness, $\gamma_1$ and $\gamma_2$ are given by
\begin{equation}
\gamma_1  = \frac{\Nt \sigma^2 }{P\taut}, \qquad \qquad
\gamma_2  = \left(1+\frac{\Nt\sigma^2}{\taut P}\right)\left(\frac{\sigma^2K\Nt\Nr}{\tauf\Pf}+\frac{\Nr\sigma^2}{\sigma^2+\taup\Pf/\Nr}\right).
\end{equation}
In essence, $\gamma_1$ captures the effect of the noise in the transmitted estimates $\widehat{\bH}^r_{i,k}$, while $\gamma_2$ captures the effect of the noise in the estimates $\widehat{\bG}_{i,k}$ as well as the noise observed during feedback. 

Having formalized the three training and analog feedback stages, we now analyze the variance,\footnote{We in fact derive the entire covariance matrix for the columns of $\bH_{i,k}-\widehat{\bH}_{i,k}$. We show that the covariance matrices are scaled identities and thus the second order statistics of the error are entirely described by a scalar variance.} $\sigmaH$, of the CSI error $\bH_{i,k}-\widehat{\bH}_{i,k}$, which automatically yields an estimated CSI variance of $1-\sigmaH$. Unfortunately, writing $\sigmaH$ exactly yields rather cumbersome expressions. For this reason, we replace the variance of the MMSE estimation error by that of a zero-forcing estimator in a manner similar to \cite{caire710multiuser, yoo2006capacity}. This ZF simplification intuitively amounts to deriving a high SNR result~\cite{caire710multiuser} and mathematically amounts to neglecting the constants $\gamma_1$ and $\gamma_2$; recall that moderately high SNR is after all the main operating region of interest for IA. Numerical results in Section \ref{sec:sims} will demonstrate that the effect of this simplification is negligible.

By neglecting $\gamma_1$ and $\gamma_2$, and after some algebraic manipulation, we find that the error $\widetilde{\bH}_{i}=\bH_{i}-\widehat{\bH}_{i}$ at the end of the three training and feedback phases can be written as 
\begin{equation}
\widetilde{\bH}_i = \sqrt{1+\frac{\Nt\sigma^2}{\taut P}}\left[\widetilde{\bH}^r_i+ \left(\widehat{\bG}_i^* \widehat{\bG}_i\right)^{-1} \widehat{\bG}_i^*\left(\sqrt{1+\frac{\Nt\sigma^2}{\taut P}}\widetilde{\bG}_i\widehat{\bH}^r_i + \sqrt{\frac{K\Nt\Nr}{\taup P_{f}}}\bV\mathbf{\Psi}_i^*\right)\right]. 
\label{eqn:error_terms}
\end{equation}
As can be seen from (\ref{eqn:error_terms}), the resulting CSI error is a combination of three terms: the first due to forward channel estimation error $\widetilde{\bH}^r_i$, the second due to feedback channel estimation error $\widetilde{\bG}_i$, and the third due to feedback noise.

To derive the statistics of $\widetilde{\bH}_i$, we note the following three facts about the three terms in (\ref{eqn:error_terms}): 
\begin{enumerate}
\item The entries of $\widetilde{\bH}^r_i$ are uncorrelated $\mathcal{CN}\left(0,\frac{\sigma^2}{\sigma^2+\frac{\taut P}{\Nt}}\right)$ variables as shown in Section \ref{sec:training}.
\item Similarly, the entries of $\widetilde{\bG}_i$ are $\mathcal{CN}\left(0,\frac{\sigma^2}{\sigma^2+\frac{\taup \Pf}{\Nr}}\right)$ implying that $\widetilde{\bG}_i\widehat{\bH}^r_i$ has independent entries with variance equal to $\frac{\Nr \sigma^2}{\sigma^2+\frac{\taup \Pf}{\Nr}}\frac{\taut P\Nt}{\sigma^2+\taut P/\Nt}$.
\item The entries of $\bV$ are uncorrelated $\mathcal{CN}(0,\sigma^2)$ variables and so are the elements of $\bV\mathbf{\Psi}_i^*$ since the matrix $\mathbf{\Psi}_i$ is unitary.  
\end{enumerate}
Combining the properties stated, the conditional covariance of each column of $\widetilde{\bH}_i$ denoted $\widetilde{\bH}_i^{(\ell)}$, conditioned of $\widehat{\bG}_i$ is~\cite{Ayach2010,Marzetta2006} 
\begin{equation}
\mathbb{E}\left(\widetilde{\bH}_i^{(\ell)}\widetilde{\bH}_i^{(\ell)*} | \widehat{\bG}_i\right)= \frac{\Nt\sigma^2}{\taut P}\bI_{\Nr}+\left(\frac{K\Nt\Nr\sigma^2}{\tauf \Pf}+\frac{\Nr\sigma^2}{\sigma^2+\taup \frac{\Pf}{\Nr}}\right)\left(\widehat{\bG}_i^*\widehat{\bG}_i\right)^{-1}.
\label{eqn:cond_cov}
\end{equation}
Since the entries of the MMSE estimate $\widehat{\bG}_i$ are Gaussian, the matrix $\left(\widehat{\bG}_i^*\widehat{\bG}_i\right)^{-1}$ has an inverse-Wishart distribution~\cite{tulino2004random}. Moreover, since $\widehat{\bG}_i$ has uncorrelated entries with a variance of $\frac{\frac{\taup \Pf}{\Nr}}{\sigma^2+\frac{\taup \Pf}{\Nr}}$, $\left(\widehat{\bG}_i^*\widehat{\bG}_i\right)^{-1}$ has a covariance matrix equal to a properly scaled identity~\cite{Ayach2010,Marzetta2006,tulino2004random}. Thus marginalizing (\ref{eqn:cond_cov}) over $\widehat{\bG}_i$, we find that $\widetilde{\bH}_i$ has independent columns with scaled identity covariance matrices with diagonal entries given by
\begin{equation}
\sigmaH= \frac{\Nt\sigma^2}{\taut P}+\frac{\sigma^2}{(K\Nt-\Nr)\Pf}\left(\frac{\Nr^2}{\taup}+\frac{K\Nt\Nr}{\tauf}\left(1+\frac{\Nr\sigma^2}{\taup \Pf}\right)\right). \label{eqn:MSE}
\end{equation}
The same high SNR simplification adopted earlier to replace MMSE estimation error by ZF estimation error, however, allows us to further simplify (\ref{eqn:MSE}) by writing
\begin{equation}
\sigmaH=\frac{\Nt\sigma^2}{\taut P} + \frac{\sigma^2}{P(K\Nt-\Nr)}\left(\frac{\Nr^2}{\gamma \taup}+\frac{K\Nt\Nr}{\gamma \tauf}\right),
\label{eqn:finalMSE}
\end{equation} 
which completes the characterization of the distortion introduced by training and analog feedback.

\emph{Note:} Finally, a word on applying the results of Section \ref{sec:IArates} to the analog feedback system described. First, we note that the analog feedback system satisfies Assumption \ref{as:decoding}, and the estimates yield $\mathbb{E}\left[\frac{P}{d} \left|(\widehat{\bw}_i^m)^*\widetilde{\bH}_{i,k}\widehat{\bff}_{k}^{\ell}\right|^2\right]=\frac{P}{d}\sigmaH$ and $\mathbb{E}\left[\frac{P}{d} \left|(\widehat{\bw}_i^m)^*\widehat{\bH}_{i,i}\widehat{\bff}_{i}^{m}\right|^2\right]=\frac{P}{d}(1-\sigmaH)$ as needed. One subtlety though is that the fading on the feedback channel introduces non-Gaussian terms into the estimates $\widehat{\bH}_{i,i}$, yet (\ref{eqn:av_rate_final}) is only exact when the estimates are truly Gaussian. For fairly accurate estimation, however,  $\widehat{\bH}_{i,i}$ can be well approximated by a Gaussian. Moreover, it will be clear from the results of Section \ref{sec:sims} that the effect of this is negligible. 

\section{Optimizing Overhead and Effective Sum-Rate} \label{sec:overhead_opt}

Having formally quantified IA sum-rate as a function of SNR and CSI quality, and characterized CSI quality in terms of training and feedback resources, we redefine both the optimization problem and objective function as
\begin{align}
\Reff^\star(P)= \max_{\taut,\taup,\tauf} \left(\frac{\Tframe -(\taut+\taup+\tauf)}{\Tframe }\right)\Rsum (\rhoeff ),
\label{eqn:opt_throughput}
\end{align}
where we have used $(\cdot)^\star$ to denote optimality. We note from (\ref{eqn:opt_throughput}) that $\rhoeff $ depends on $\sigma^2_{\widetilde{\bH}}$ and thus on $\taut$, $\taup$, and $\tauf$. The problem in (\ref{eqn:opt_throughput}) can be rewritten in a more tractable form as~\cite{kobayashi2009training}
\begin{align}
\Reff^\star(P)= \max_{\substack{\alpha \\ \alpha_\mathrm{min}\leq\alpha\leq 1}} \left[\left(1-\alpha\right)\max_{\substack{\taut,\ \taup,\ \tauf \\ \taut+\taup+\tauf= \alpha \Tframe }}\Rsum (\rhoeff )\right],
\label{eqn:opt_throughput2}
\end{align}
where $\alpha_\mathrm{min}=K(\Nt+\Nr+K\Nt)/\Tframe$ and is dictated by the minimum number of training and feedback symbols needed to render the estimation problems in Section \ref{sec:analog_feedback} well defined. The inner maximization in (\ref{eqn:opt_throughput2}) optimizes sum-rate for a fixed overhead length of $\Tohd =\alpha \Tframe $ and the outer maximization finds the optimal $\alpha$ thereby completing the solution.

Since $\Rsum (\rhoeff )$ is decreasing in $\sigma^2_{\widetilde{\bH}}$, the inner maximization step simplifies to
\begin{align}
\begin{split}
\sigma^{2\star}_{\widetilde{\bH}} =\min_{\taut,\ \taup,\ \tauf}\hspace{8pt} &  \frac{\Nt\sigma^2}{\taut P} + \frac{\sigma^2}{P(K\Nt-\Nr)}\left(\frac{K\Nt\Nr}{\gamma \tauf}+\frac{\Nr^2}{\gamma \taup}\right) \\ &
s.t.\qquad \taut+\taup +\tauf=\alpha \Tframe .
\label{eqn:inner_max}
\end{split}
\end{align}
Although (\ref{eqn:inner_max}) is an integer problem, its continuous relaxation is convex. Applying standard convex optimization techniques, the Lagrangian for the inner maximization is
\small
\begin{align}
\begin{split}
\Lambda(\taut,\taup,\tauf,\lambda)=\frac{\Nt\sigma^2}{\taut P} + \frac{\sigma^2}{P(K\Nt-\Nr)}\left(\frac{K\Nt\Nr}{\gamma \tauf}+\frac{\Nr^2}{\gamma \taup}\right) +\lambda\left(\taut+\taup+\tauf-\alpha \Tframe\right).
\end{split}
\end{align}
\normalsize
Solving for the first order KKT conditions, we obtain the optimal training and feedback times as a function of the total overhead budget $\alpha \Tframe$ as
\begin{equation}
\taut^\star=\frac{\sqrt{\gamma\Nt(K\Nt-\Nr)}}{\mu}\alpha \Tframe, \quad\qquad
\taup^\star=\frac{\Nr}{\mu}\alpha \Tframe, \quad\qquad
\tauf^\star=\frac{\sqrt{K\Nt\Nr}}{\mu}\alpha \Tframe,\nonumber
\end{equation}
where $\mu=\sqrt{\gamma\Nt(K\Nt-\Nr)}+\Nr+\sqrt{K\Nt\Nr}$. After solving the problem's continuous relaxation, convexity implies that for any given feasible overhead budget $\alpha\Tframe$ simply examining the few integer neighbors of the points $\taut^\star$, $\taup^\star$, $\tauf^\star$ yields the integer training and feedback times that minimize CSI distortion, i.e., optimal integer training and feedback times can be found by a simple search over the grid neighbors of the non-integer solution. Proceeding with the continuous relaxation, the minimum CSI distortion for an overhead budget $\alpha \Tframe$ is
\begin{equation}
\sigma^{2\star}_{\widetilde{\bH}} = \frac{\sigma^2 \left(\sqrt{K\Nt\Nr}+\Nr+\sqrt{\gamma \Nt(K\Nt-\Nr)}\right)^2}{\gamma P (K\Nt-\Nr)\alpha \Tframe }.
\label{eqn:optimalMSE}
\end{equation}

Having found the optimal allocation of $\taut$, $\taup$, and $\tauf$ for a fixed overhead budget, what remains is to optimize the budget itself. The outer optimization in (\ref{eqn:opt_throughput}), however, does not admit a closed form solution. To circumvent this problem, prior work on single user and broadcast channels has specialized their results to the limiting high or low SNR regimes~\cite{lozano2008interplay, hassibi_training}, relied on numerical optimization~\cite{Ayach2010}, or resorted to characterizing the scaling of overhead with various system parameters based on sum-rate lower bounds~\cite{kobayashi2009training}. To give accurate results on finite-SNR sum-rate, we propose to optimize a series expansion of (\ref{eqn:throughput_simple}) with respect to the channel's Doppler spread around the point $\fd=0$~\cite{jindal2010unified}. Recall that $\Tframe$ which we have been using thus far is related to $\fd$ by the relationship $\Tframe=\frac{1}{2\fd}$. To that end, we give the following result on the series expansion of $\Reff(P, \Tohd )$.

\begin{proposition}
The effective sum-rate achieved by IA with training and feedback expands as
\begin{align}
\begin{split}
 \Reff(P, \Tohd ) = &  (1- \alpha)(1 + \rho Kd)\left[\frac{\Rsum (\rho)}{1+\rho Kd}  -\frac{2\beta}{d\alpha}\dRsum (\rho)\fd \right. \\  &\hspace{36pt} \left.  +  \left(\frac{2\beta}{d\alpha}\right)^2\left(\ddRsum (\rho)(1+\rho Kd)+2Kd\dRsum (\rho)\right)\frac{\fd^2}{2}\right] +O(\fd^3), 
\end{split}
\end{align}
where 
\begin{equation}
\beta=\frac{\left(\sqrt{K\Nt\Nr}+\Nr+\sqrt{\gamma \Nt(K\Nt-\Nr)}\right)^2}{\gamma(K\Nt-\Nr)},
\label{eqn:beta}
\end{equation} 
whereas $\dRsum (\rho)$ and $\ddRsum (\rho)$ are the first and second derivatives of \emph{perfect CSI sum-rate}, $\Rsum (\rho)$, which can be conveniently expressed as 
\small
\begin{align}
\dRsum (\rho) & =\frac{1}{\rho}\left(Kd\log_2(e)-\frac{\Rsum (\rho)}{\rho}\right), \qquad
\ddRsum (\rho) & = -\frac{1}{\rho^2}\left(Kd\log_2(e)+\dRsum (\rho)-2\frac{\Rsum (\rho)}{\rho}\right).
\end{align}
\normalsize
\label{res:expansion}
\end{proposition}
\begin{IEEEproof}
Given in Appendix \ref{sec:proof_expansion}.
\end{IEEEproof}

Thus, by expanding effective sum-rate w.r.t. $\fd$, we have transformed the complicated non-linear dependence of effective sum-rate on system parameters such as $P$, $\Tframe $, $\fd$, and $\Tohd $ to a simpler polynomial dependence. The expansion in Proposition \ref{res:expansion} can now be used to derive the expansion of the optimal overhead budget, $\alpha^\star$, along with the performance it achieves. Relaxing the constraint that the overhead fraction $\alpha$ must be rational, simply differentiating the series expansion in Proposition \ref{res:expansion} and equating it to zero yields the optimal overhead budget $\alpha^\star$.
\begin{proposition}
The optimum overhead fraction $\alpha^\star$ for an IA system with training and analog feedback expands as
\begin{align}
\begin{split}
\alpha^\star & =\sqrt{\frac{2\beta(1+\rho Kd)}{d}\frac{\dRsum (\rho)}{\Rsum (\rho)}\fd} - \frac{\beta}{d}\left(\frac{\ddRsum (\rho)}{\dRsum (\rho)}(1+\rho Kd)+2Kd\right)\fd+O(\fd^{3/2}), 
\end{split}
\end{align}
which results in the optimal effective sum-rate
\begin{align}
\begin{split}
\Reff^\star(P)= & \Rsum (\rho)-2\sqrt{\frac{2\beta}{d}(1+\rho Kd)\dRsum (\rho)\Rsum (\rho)\fd} +O(\fd).
\end{split}
\end{align}
Note that if $\fd$ is large enough that $\alpha^\star<\alpha_\mathrm{min}$ the optimal overhead budget must be adjusted to $\alpha_\mathrm{min}$ and the expression for $\Reff^\star(P)$ correspondingly updated. 
\label{res:end_result}
\end{proposition}
\begin{IEEEproof}
The proof follows directly from differentiating the expansion in Proposition \ref{res:expansion} w.r.t. $\alpha$ and solving the resulting cubic polynomial for its relevant root.
\end{IEEEproof}

Therefore, Proposition \ref{res:end_result} along with the solution to (\ref{eqn:inner_max}) gives the effective sum-rate-maximizing amount of forward training, feedback channel training, and analog feedback as simple functions of fundamental system parameters such as SNR, Doppler spread (equivalently $\Tframe $), and perfect CSI sum-rate. Numerical results in Section \ref{sec:sims} will show that the overhead expansion in Proposition \ref{res:end_result} is accurate for a wide range of system parameters and can thus obviate the need for numerical overhead optimization. Furthermore, the derived results allow us to draw several interesting insights into IA system design and performance:
\begin{enumerate}
\item  The optimal overhead budget $\alpha$ scales with $\sqrt{\fd}$. As stated, for high enough Doppler $\alpha^\star$, must be adjusted to $\alpha_\mathrm{min}$ meaning that overhead subsequently increases with $\fd$. This scaling behavior is in line with previous results on other single and multiuser channels. 
\item The sum-rate penalty due to overhead and imperfect CSI behaves similarly, i.e., increases with $\sqrt{\fd}$ initially and with $\fd$ at high Doppler. 
\item Examining the leading term in $\alpha^\star$ we note that, similarly to ~\cite{jindal2010unified}, the term $(1+\rho Kd)\frac{\dRsum (\rho)}{\Rsum (\rho)}$ behaves like $Kd/\log_e(1+\rho)$ and thus the optimal overhead budget decreases with SNR roughly as $\sqrt{Kd/\log_e(1+\rho)}$.
\item Since overhead decreases with SNR, a minimum overhead interval of $K\Nt+K\Nr+K^2\Nt$ is always optimal at sufficiently high SNR. Thus, the effective number of spatial DoF achieved by IA with the analog feedback strategy described is $\left(1-\frac{K\Nt+K\Nr+K^2\Nt\Nr}{\Tframe}\right)Kd$, i.e., the  DoF penalty increases linearly with $\fd$. 
\item Again examining the leading term in $\alpha^\star$ we note that it increases with $\sqrt{\beta}$. Recalling the definition of $\beta$ in (\ref{eqn:beta}), we conclude that the optimal overhead budget increases with $\sqrt{P/\Pf}$. This formalizes the relationship between overhead and feedback link quality.
\end{enumerate}
In addition to highlighting the dependence of overhead and effective sum-rate on various system parameters, the derived results can provide simple answers to various network design questions. For example, by simply comparing IA's effective sum-rate expression to those achieved by other transmission strategies, one can choose the optimal transmission strategy for a given fading environment. Moreover, since overhead and channel selectivity have been shown to place fundamental limits on the gains of cooperation in wireless network~\cite{Lozano-Heath-Andrews}, the overhead-aware analysis presented in this paper can help in determining the optimal number of cooperative IA users at a given level of selectivity. 

Consider, as a simple example, a $K$-user single-stream cooperation cluster with a variable number of antennas in which extra users are allowed to cooperate via IA if they do not incur a loss in effective sum-rate, else the extra users are not allowed access to the propagation medium and presumably left to transmit on a separate channel. In this model, additional cooperating users can be incorporated into the cluster as long as $\mathcal{I^\star}_{K+1}(P)-\mathcal{I^\star}_{K}(P)>0$ where we have made cluster size explicit in the effective sum-rate subscript. Consequently, the effective sum-rate-maximizing cluster size becomes the smallest $K$ such that $\mathcal{I^\star}_{K+1}(P)-\mathcal{I^\star}_{K}(P)<0$. Moreover, note that minimizing overhead and maintaining IA feasibility imposes the constraint $\Nt+\Nr= K+1$~\cite{razaviyayn_DoF}. Thus, writing $\Nt$ and $\Nr$ in terms of $K$, e.g. $\Nt=\lceil(K+1)/2\rceil$, the user admission rule can be simplified to a function of only $K$, $\fd$, SNR, and $\gamma$. To simplify the user admission rule even further, we make the following approximations: (i) we consider the leading term in $\Reff(P,\Tohd)$ thus focusing on IA's effective DoF 
given in the fourth observation after Proposition \ref{res:end_result}, (ii) we assume that $\Nt=(K+1)/2$ and thus relax its integer constraint. Using these simplifications, the user admission rule $\mathcal{I^\star}_{K+1}(P)-\mathcal{I^\star}_{K}(P)>0$ simplifies to
\begin{equation}
4K^3+15K^2+17K+6<\frac{1}{\fd},
\label{eqn:admission}
\end{equation}
i.e., a $K$-user cluster can be extended to $K+1$ as long as (\ref{eqn:admission}) is satisfied. Interestingly, this implies that in such single-stream IA scenarios the effective sum-rate-maximizing cluster size grows with $\fd^{-1/3}$. While the approximate admission rule is a rather simplified version of $\mathcal{I^\star}_{K+1}(P)-\mathcal{I^\star}_{K}(P)>0$, we show in Section \ref{sec:sims} that it is very accurate at predicting optimal cluster size. Finally, we note that while we provide this example to illustrate problems that can be solved using our analysis, the rule in (\ref{eqn:admission}) is by no means universal. When parameters such as large-scale fading or uncoordinated interference are considered, both the analysis and the admission rule must be adjusted.


\section{Simulation Results} \label{sec:sims}

\begin{figure}[t!]
 \centering
	\includegraphics[width=4in]{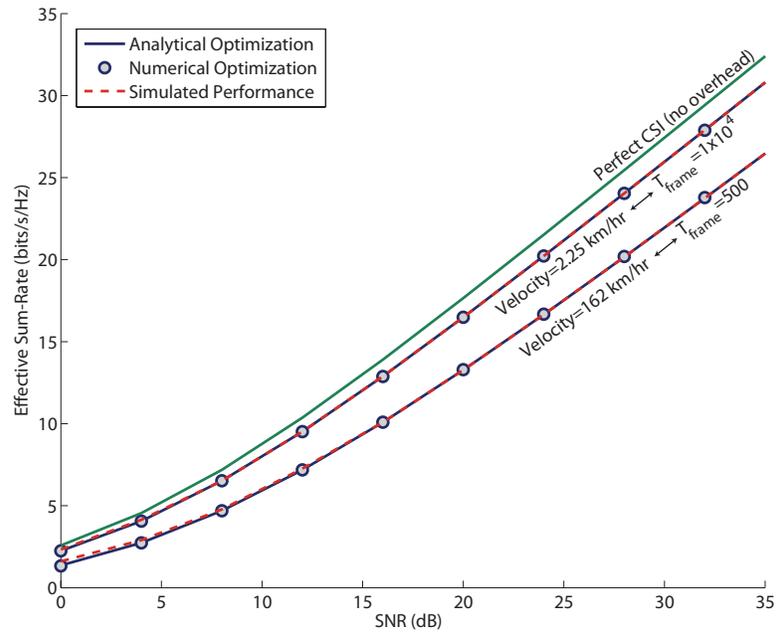}
	\caption{Effective Sum-Rate vs. SNR for systems with different normalized Doppler spreads. This quantifies the loss in sum-rate due to both imperfect CSI and overhead and shows that the performance predicted by the analytical results presented is an accurate representation of optimal performance.}
	\label{fig:throughput_vs_SNR}
\vspace{-9pt}
\end{figure}
\begin{figure}[t!]
 \centering
	\includegraphics[width=4in]{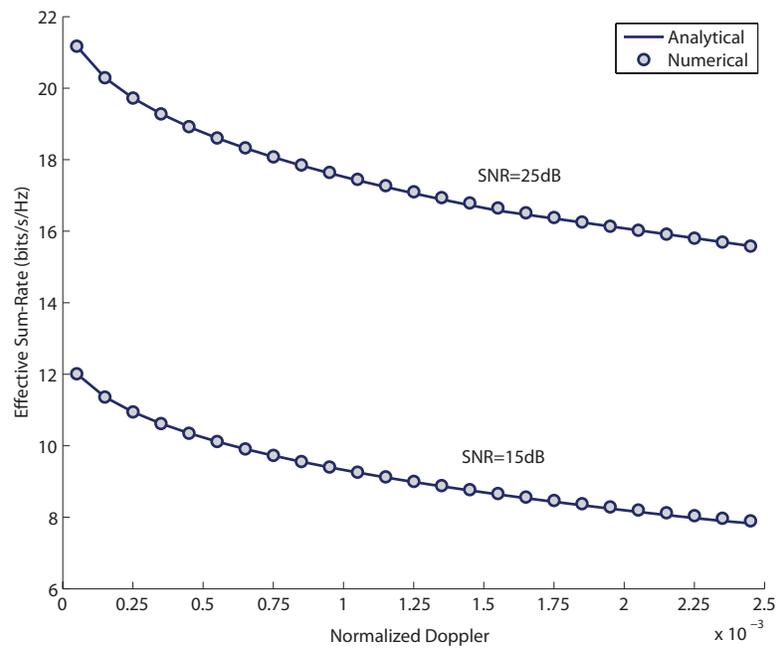}
	\caption{Effective Sum-Rate vs. Normalized Doppler for IA systems at different SNR levels. This quantifies the degradation in sum-rate as mobility increases resulting in an increased overhead penalty.}
	\label{fig:throughput_vs_doppler}
\vspace{-9pt}
\end{figure}

Consider a three-user IA cluster with two transmit antennas, two receive antennas, and one spatial stream per user and let $\gamma=\frac{\Pf}{P}=1$. Fig. \ref{fig:throughput_vs_SNR} shows the effective sum-rate achieved by IA in systems with various levels of mobility or normalized Doppler spreads, $\fd$. To quantify the degradation in effective sum-rate caused by overhead and imperfect CSI, we include the performance of a baseline genie-aided system in which CSI is both perfect and free. Fig. \ref{fig:throughput_vs_SNR} indicates that IA achieves good performance in a system with vehicular-levels of mobility. In fact, if typical wireless parameters are adopted, such as a wavelength of $\lambda=0.15\ m$ (corresponding to a carrier frequency of 2 GHz), a coherence bandwidth of $W_\mathrm{C}=300\ kHz$, and a normalized Doppler given by $\fd=\frac{v}{\lambda W_\mathrm{C}}$ where $v$ is the user's velocity, Fig. \ref{fig:throughput_vs_SNR} indicates that IA could theoretically perform well even at a speed of more than $160\ km/hr$. The rate of performance degradation over a wider range of Doppler spread can be seen in Fig. \ref{fig:throughput_vs_doppler}. Both Figs. \ref{fig:throughput_vs_SNR} and \ref{fig:throughput_vs_doppler} indicate that the analytical results of Section \ref{sec:overhead_opt} are very effective in optimizing the effective sum-rate of IA systems as the resulting performance closely matches that of a numerically optimized system. Finally, Fig. \ref{fig:throughput_vs_SNR} indicates that the effect of the simplifying assumptions made in Section \ref{sec:analog_feedback} is negligible since the effective sum-rate predicted by the derived rate expressions closely matches simulated IA performance. A very slight deviation is noticed at very low SNR where the ZF simplification in Section \ref{sec:analog_feedback} is a less accurate approximation of MMSE performance.

\begin{figure}[t!]
 \centering
	\includegraphics[width=4in]{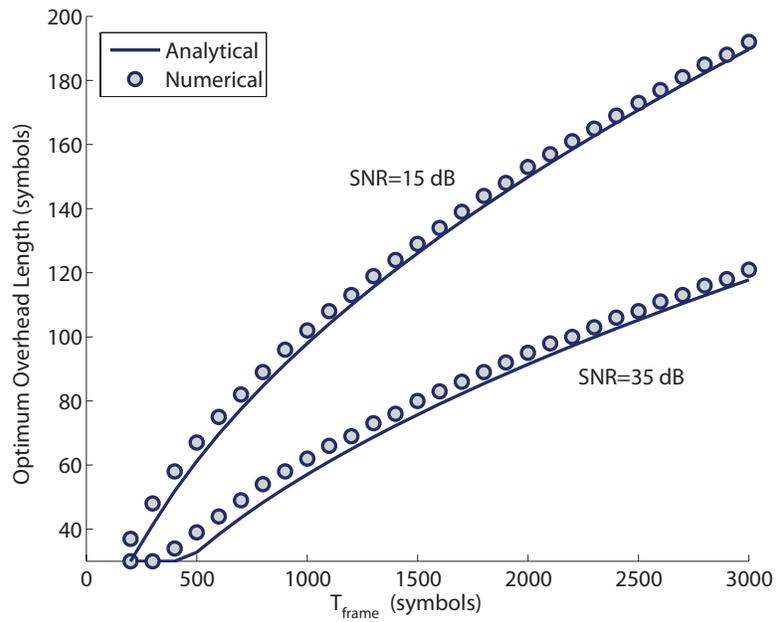}
	\caption{$\Tohd $ vs. $\Tframe $. This confirms that the optimal value of $\Tohd $ scales with $\sqrt{\Tframe }$ as predicted, and shows that optimizing a series expansion of the objective yields remarkably accurate results.}
	\label{fig:overhead_vs_Tframe}
\end{figure}

\begin{figure}[t!]
 \centering
	\includegraphics[width=4in]{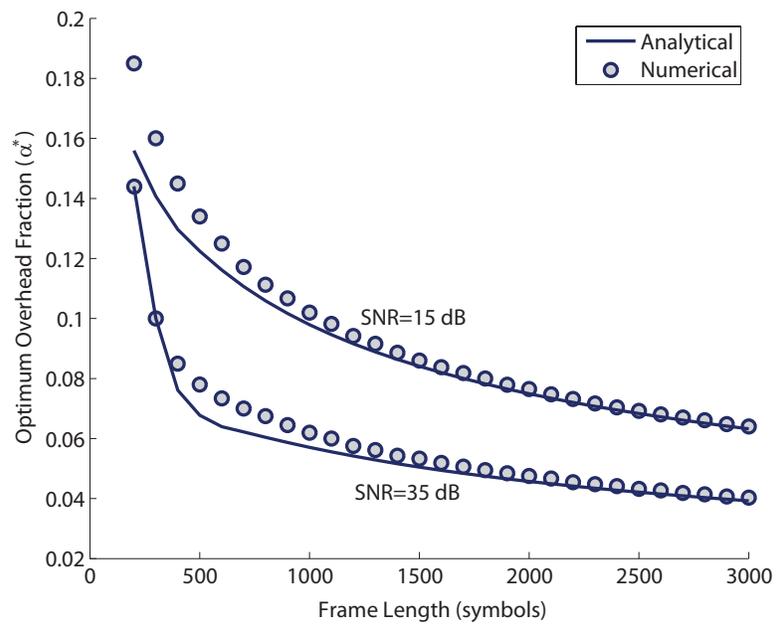}
	\caption{$\alpha^\star $ vs. $\Tframe $. This confirms that the optimal value of $\alpha $ scales with $\frac{1}{\sqrt{\Tframe }}$ and thus scales with $\sqrt{\fd}$ as predicted.}
	\label{fig:overheadfraction_vs_Tframe}
\end{figure}

\begin{figure}[t!]
 \centering
	\includegraphics[width=4in]{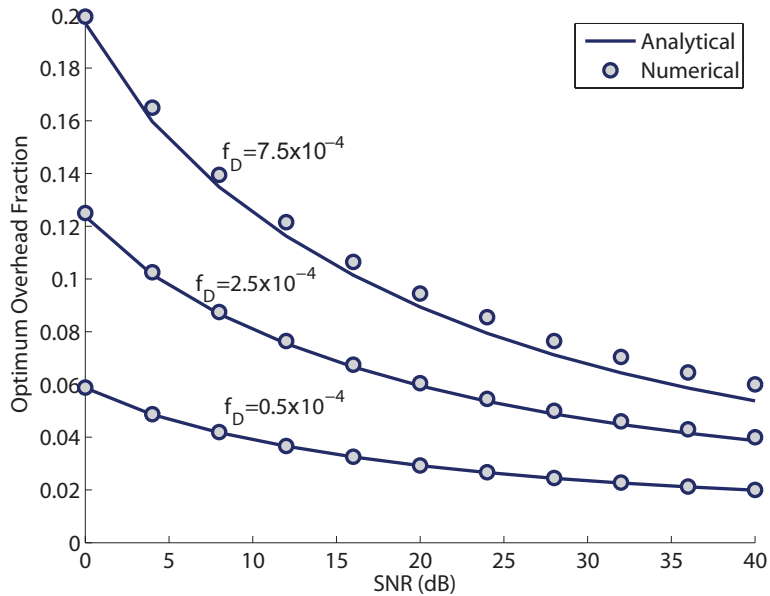}
	\caption{$\alpha^\star $ vs. SNR. This shows the decrease of the optimal overhead budget with SNR. As stated in Section \ref{sec:overhead_opt}, it can be shown that the decrease is logarithmic with SNR. The figure also demonstrates that our expansion-base results are very accurate, deviating only slightly in high-SNR high-mobility scenarios.}
	\label{fig:overhead_vs_SNR}
\end{figure}

\begin{figure}[t!]
\centering
\subfigure[$\alpha^\star $ vs. $1/\gamma$]{
\includegraphics[width=3.1in]{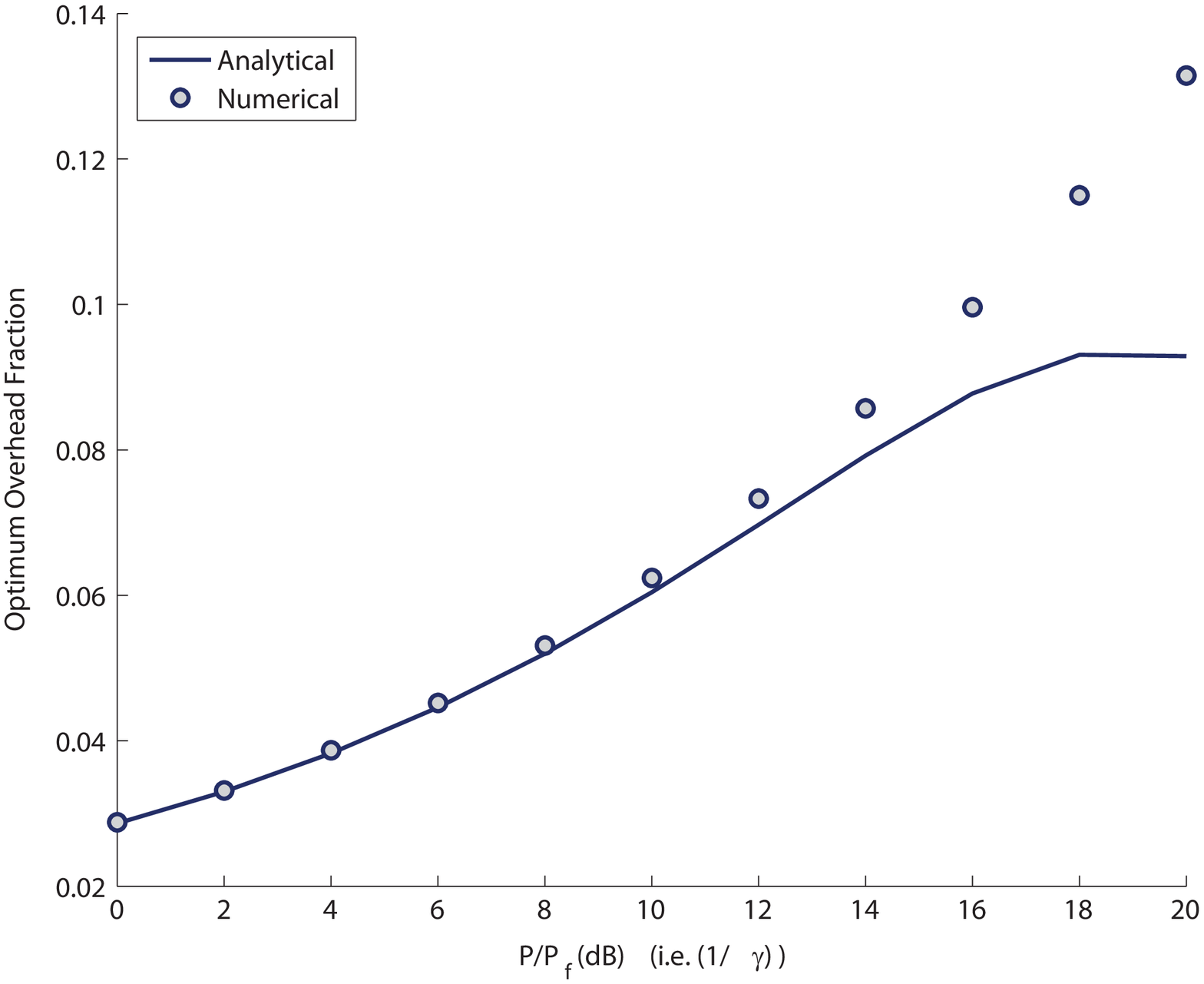}
\label{fig:alpha_vs_gamma}
}
\subfigure[Effective Sum-Rate vs. $1/\gamma$]{
\includegraphics[width=3.1in]{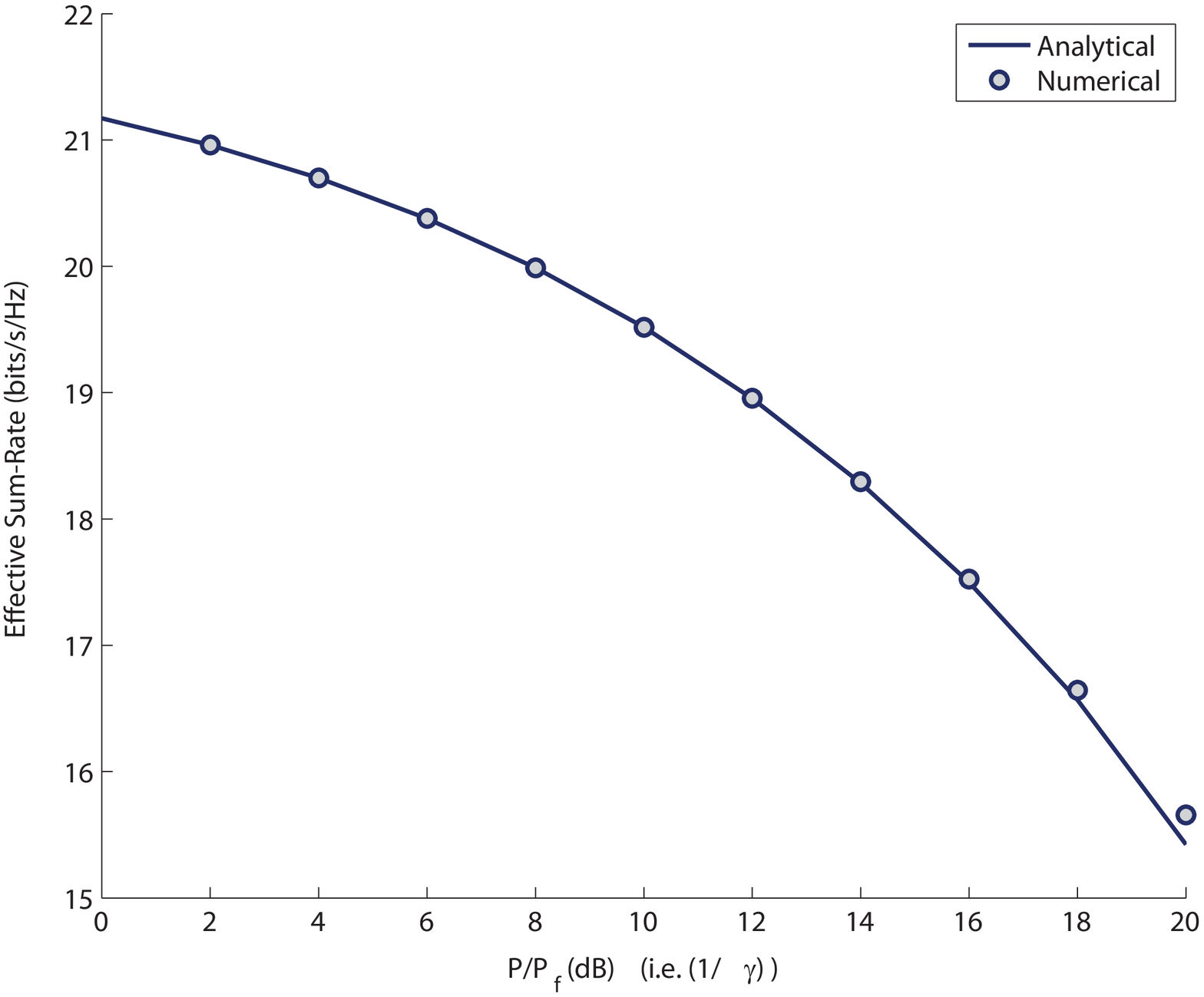}
\label{fig:throughput_vs_gamma}
}
\caption{This figure shows the relationship between $\alpha^\star$ and $\Reff^\star (P)$ with the feedback channel's relative quality for a system with $\Tframe=10^4$. Plot (a) verifies the increase of overhead with $1/\gamma$, when plot in linear scale the square root rate of increase can be verified. Plot (b) verifies the rate of decrease of optimal effective sum-rate with feedback link quality.}
\label{fig:effect_of_gamma}
\end{figure}

Fig. \ref{fig:overhead_vs_Tframe} shows the optimal overhead budget for systems with varying frame lengths and again includes both the analytical overhead budget from Section \ref{sec:overhead_opt} as well as the result of numerically optimizing the same system. Fig. \ref{fig:overhead_vs_Tframe} confirms that  $\Tohd $ increases with frame size $\Tframe$ at a rate proportional to $\sqrt{\Tframe}$. Thus $\alpha^\star$ indeed decreases with $\frac{1}{\sqrt{\Tframe}}$, as shown in Fig. \ref{fig:overheadfraction_vs_Tframe}, or equivalently increases with $\sqrt{\fd}$ (and with $\fd$ for sufficiently high Doppler). Fig. \ref{fig:overhead_vs_Tframe} also shows that the expansion in Proposition \ref{res:end_result} provides an accurate characterization of IA's effective sum-rate-maximizing overhead budget over a wide range of SNRs and frame sizes. Fig. \ref{fig:overhead_vs_SNR} in turn verifies the decrease of $\alpha^\star$ with SNR, which as stated in Section \ref{sec:overhead_opt} follows the relationship $\alpha^\star\sim\sqrt{\log_e(1+\rho)^{-1}}$. To complete the characterization of overhead and effective sum-rate, Fig. \ref{fig:effect_of_gamma} quantifies the deleterious effect of a weak feedback channel on overhead and effective sum-rate. Fig. \ref{fig:effect_of_gamma} also indicates that the expansion results of Section \ref{sec:overhead_opt} could significantly underestimate $\alpha^\star$ in very weak feedback channels, though the final effect on throughput remains limited.

\begin{figure}[t!]
 \centering
	\includegraphics[width=4in]{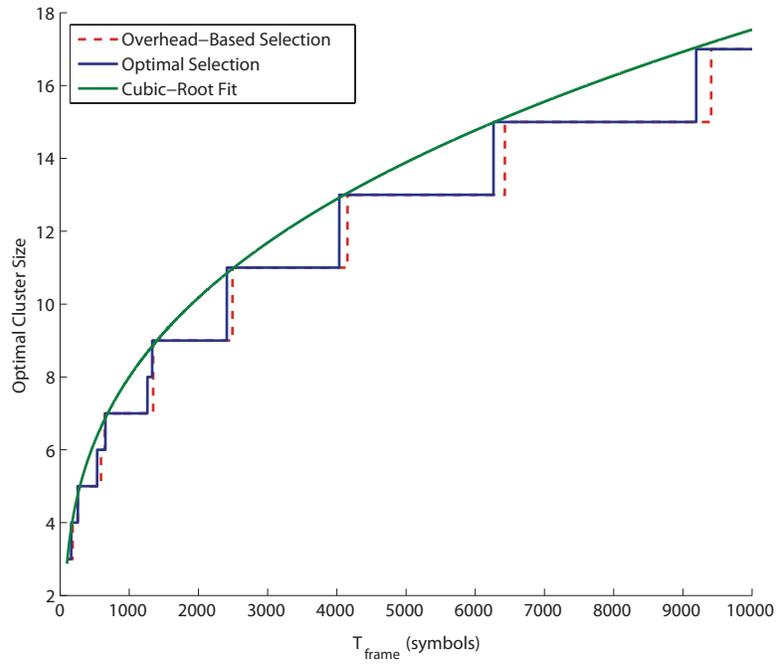}
	\caption{Optimal Cluster Size vs. $\Tframe$. This shows the optimal number of users to coordinate via IA which increases channels coherence time. This also shows that comparing overhead, i.e., overhead based selection, provides accurate decisions on optimal cluster size.}
	\label{fig:network_size_vs_Tframe}
\end{figure}

\begin{figure}[t!]
 \centering
	\includegraphics[width=4in]{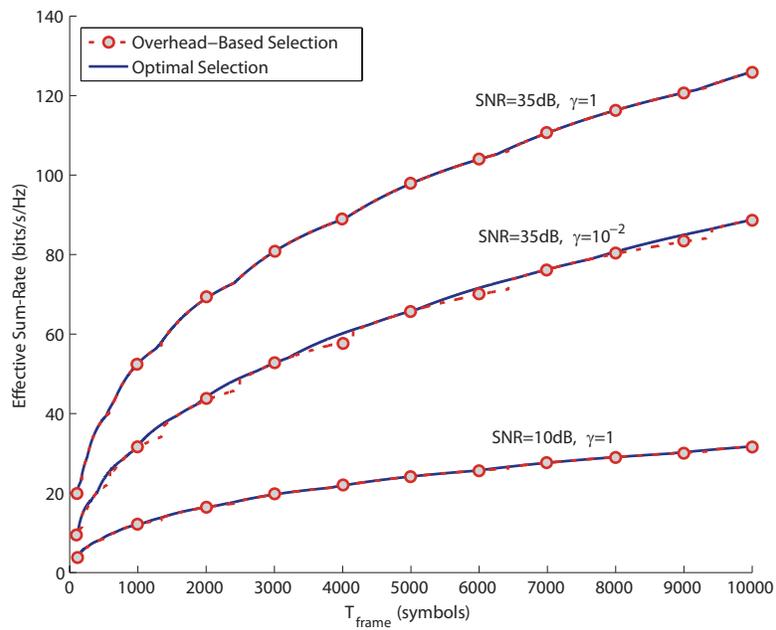}
	\caption{Effective Sum-Rate with Cluster Size Optimization vs. $\Tframe$. This shows the increase in effective sum-rate as a function of $\Tframe$ when the cluster size is chosen to maximize rate. This also quantifies the minimal sum-rate loss due to sub-optimal overhead-only based cluster sizing.}
	\label{fig:User_selection_throughput_vs_Tframe}
\end{figure}

Finally we examine the efficiency of our overhead analysis in further network design. We consider the motivating example given in Section \ref{sec:overhead_opt} of a $K$-user system for which we seek to optimize the cooperation cluster size as a function of mobility. Fig. \ref{fig:network_size_vs_Tframe} shows the optimal cluster size as a function of $\Tframe$ for an IA system at 35 dB SNR. Fig \ref{fig:User_selection_throughput_vs_Tframe} shows the corresponding effective sum-rate achieved. We plot the cluster size and effective sum-rate resulting from (i) an exhaustive search over all possible $K$, and (ii) the simple overhead-based user admission rule in (\ref{eqn:admission}). We note that the cluster size predicted by the two methods are in close agreement, and that the asymptotic cube-root relationship predicted in Section \ref{sec:overhead_opt} between optimal cluster size and $\Tframe$ is quite accurate even for small values of $K$. While the overhead-based rule tends to underestimate cluster size for small intervals of $\Tframe$, Fig. \ref{fig:User_selection_throughput_vs_Tframe} indicates that the resulting rate gap from optimal sizing is negligible. The same can be said about the rate loss when applying the same overhead based rule to a system at an SNR of 10 dB and a system with $\gamma=10^{-2}$.


\section{Conclusion} \label{sec:Conclusion}

We considered IA's effective sum-rate in practical systems where CSI is imperfect and comes with an associated overhead cost. We showed that training and feedback overhead can be optimized to ensure good IA performance over a wide range of SNR and Doppler spread. We quantified the dependence between overhead and various system parameters such as feedback link quality. More sophisticated precoding algorithms, designed to be robust to imperfect CSI, could further improve the demonstrated performance and thus remain a promising area for future work. The derived results provide a formal method to gauge true IA performance vs. other transmission strategies, and can thus highlight settings under which IA provides tangible gains. The derived analysis can also be used for further network design as demonstrated by the motivating example given at the end of Section \ref{sec:overhead_opt} on overhead-aware user admission and optimal network sizing. 

\appendices 
\section{Proof of Proposition \ref{res:expansion}} \label{sec:proof_expansion}

To expand effective sum-rate around $\fd=0$, we start computing its first order derivative
\small
\begin{equation}
\frac{\partial \Reff(P,\Tohd)}{\partial \fd}=(1-\alpha)\dRsum(\rho)\frac{\partial\rhoeff}{\partial\fd}|_{\fd=0}=-(1-\alpha)\dRsum(\rho)(1+Kd\rho)\frac{2\beta}{d\alpha}
\label{eqn:firstderiv}
\end{equation}
\normalsize
where $\frac{\partial\rhoeff}{\partial\fd}|_{\fd=0}$ is evaluated by noticing that after solving the inner maximization in (\ref{eqn:inner_max}) and obtaining ${\sigmaH}^\star$ in (\ref{eqn:optimalMSE}) we have $\frac{\partial\sigmaH}{\partial\fd}|_{\fd=0}=\frac{2\beta}{d\alpha}$. The term $\dRsum(\rho)$ can be obtained by a standard derivation of the exponential integral rate expression in (\ref{eqn:av_rate}) w.r.t $\rho$ and is given directly in the statement of Proposition \ref{res:expansion}; $\ddRsum(\rho)$ is obtained similarly.  As for the second order term, we have
\small
\begin{align}
\frac{\partial^2 \Reff(P,\Tohd)}{\partial \fd^2} & = (1-\alpha)\left[\ddRsum(\rho)\left(\frac{\partial\rhoeff}{\partial\fd}\right)^2+\dRsum(\rho)\frac{\partial^2\rhoeff}{\partial\fd^2}\right]|_{\fd=0} \nonumber \\
& \stackrel{(a)}{=} (1-\alpha)\left[\ddRsum(\rho)\left(\frac{\partial\rhoeff}{\partial\fd}\right)^2+ \dRsum(\rho)\left(\frac{\partial^2\rhoeff}{\partial{{\sigmaH}^\star}^2 }\left(\frac{\partial{\sigmaH}^\star}{\partial\fd}\right)^2+\frac{\partial\rhoeff}{\partial{\sigmaH}^\star}\frac{\partial^2{\sigmaH}^\star}{\partial\fd^2}\right)\right]|_{\fd=0}
\nonumber \\
& \stackrel{(b)}{=}  (1-\alpha)\left(\frac{2\beta}{d\alpha}\right)^2(1+\rho Kd)\left(\ddRsum(\rho)(1+\rho Kd)+2Kd\dRsum(\rho)\right)
\label{eqn:secondderiv}
\end{align}
\normalsize
where $(a)$ expands $\frac{\partial^2\rhoeff}{\partial\fd^2}$ for clarity and $(b)$ is by noticing that $\frac{\partial^2{\sigmaH}^\star}{\partial\fd^2}=0$ since ${\sigmaH}^\star$ is linear in $\fd$ and otherwise replacing the values of the different variables. Combining (\ref{eqn:firstderiv}) and (\ref{eqn:secondderiv}) we get the resulting second order expansion. Higher order expansions can be found if additional accuracy is needed, however, the second order expansion is in general sufficient. 

\singlespace
\bibliographystyle{IEEEtran}
\bibliography{IEEEabrv,ia_overhead}

\newpage

\end{document}